\documentclass[]{elsarticle}

\usepackage{hyperref}

\journal{Journal of Magnetism and Magnetic Materials}

\begin{document}

\begin{frontmatter}

\title{Penetration depth measurements of K$_2$Cr$_3$As$_3$ and Rb$_2$Cr$_3$As$_3$}

\author[mymainaddress]{G. M. Pang}
\author[mymainaddress]{M. Smidman}
\author[mymainaddress]{W. B. Jiang}
\author[Beijing]{Y. G. Shi}
\author[mymainaddress]{J. K. Bao}
\author[mymainaddress]{Z. T. Tang}
\author[mymainaddress]{Z. F.  Weng}
\author[mymainaddress]{Y. F. Wang}
\author[mymainaddress]{L. Jiao}
\author[mymainaddress]{J. L. Zhang}
\author[Beijing]{J. L. Luo}
\author[mymainaddress,mysecondaryaddress]{G. H. Cao}
\author[mymainaddress,mysecondaryaddress]{H. Q. Yuan\corref{mycorrespondingauthor}}
\cortext[mycorrespondingauthor]{Corresponding author}
\ead{hqyuan@zju.edu.cn}

\address[mymainaddress]{Center for Correlated Matter and Department of Physics, Zhejiang University, Hangzhou 310058, China}
\address[mysecondaryaddress]{Collaborative Innovation Center of Advanced Microstructures, Nanjing 210093, China}
\address[Beijing]{Beijing National Laboratory for Condensed Matter Physics and Institute of Physics, Chinese Academy of Science, Beijing 100190, China}

\begin{abstract}

The newly discovered superconductors $A_2$Cr$_3$As$_3$ ($A$~=~K, Rb, Cs), with a quasi-one-dimensional crystal structure have attracted considerable interest. The crystal structure consists of double-walled tubes of [Cr$_3$As$_3$]$^{2-}$ that extend  along the $c$ axis. Previously we reported measurements of the change in London penetration depth of polycrystalline samples of  K$_2$Cr$_3$As$_3$ using a tunnel diode oscillator based technique, which show a linear temperature dependence at low temperatures, giving evidence for line nodes in the superconducting gap. Here we report similar measurements of the penetration depth for polycrystalline Rb$_2$Cr$_3$As$_3$ and several single crystals of  K$_2$Cr$_3$As$_3$, prepared by two different research groups.  The single crystal measurements show similar behavior to polycrystalline samples down to 0.9-1.2~K, where a downturn is observed in the frequency shift for all single crystal samples. These results give further evidence for nodal superconductivity in  K$_2$Cr$_3$As$_3$, which indicates that the superconducting pairing state is unconventional. The different low temperature behavior observed in samples which have deteriorated after being exposed to air, emphasises that it is necessary to properly handle the samples prior to being measured because the $A_2$Cr$_3$As$_3$ compounds are extremely air sensitive and evidence for nodal superconductivity from penetration depth measurements is only observed in the samples which display a sharp superconducting transition.  Therefore further work is required to improve the quality of single crystals and to identify the origin of the downturn.
\end{abstract}


\end{frontmatter}


\section{Introduction}

\begin{figure}[tb]
\begin{center}
  \includegraphics[width=0.75\columnwidth]{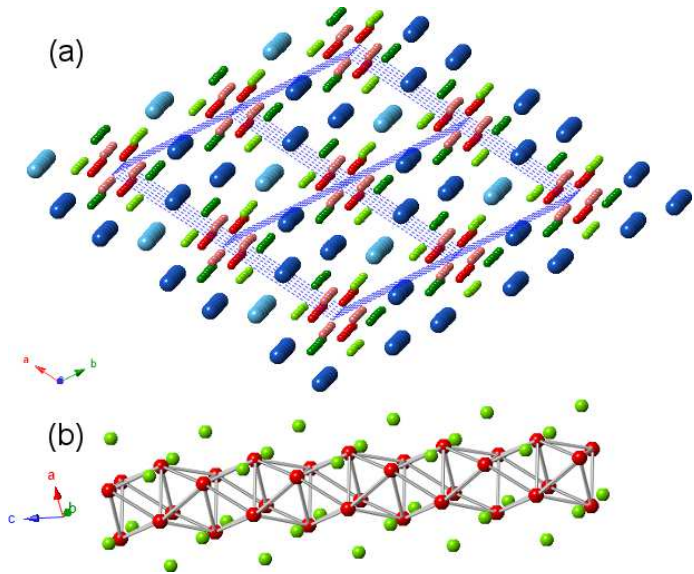}
\end{center}
	\caption{(a) Crystal structure of  $A_2$Cr$_3$As$_3$ ($A$~=~K, Rb, Cs), with the Cr atoms shown in red, As in green and $A$ in blue. The two different colors for each element indicate the two inequivalent sites. The dashed lines mark the boundaries of the unit cell and the double walled nanotubes are located at the corners of four cells. (b) A sideways view of one of the nano-tubes, showing both the Cr atoms on the inside and the As on the outside. The solid lines show how the atoms form face-sharing octahedra.}
   \label{figure1}
\end{figure}

Superconductivity in Cr-based compounds has recently attracted attenton after the discovery that CrAs becomes superconducting with $T_c\approx2~$K under an applied pressure of $p_c\approx0.8~$GPa, where antiferromagnetism is suppressed \cite{Wu2014superconductivity}. Subsequently, a family of related superconductors $A_2$Cr$_3$As$_3$ ($A$~=~K, Rb, Cs) were also discovered, with a highest transition temperature of $T_c=6.1~$K for K$_2$Cr$_3$As$_3$\cite{Bao2015superconductivity,Tang2015unconventional,Tang2015superconductivity}. The crystal structure of  $A_2$Cr$_3$As$_3$ shown in Fig.~\ref{figure1} displays quasi-one-dimensional (q1D) structural features, namely [(Cr$_3$As$_3$)$^{2-}]^\infty$ double-walled tubes which extend along the $c$-axis, separated by alkaline metal cations. The structure of the tubes is shown in Fig.~\ref{figure1}(b), where the inner tube consists of Cr atoms, surrounded by a tube of As atoms. Evidence for unconventional superconductivity in K$_2$Cr$_3$As$_3$ arises from the absence of a Hebel Slichter coherence peak in NMR measurements \cite{K2Cr3As3NMR} and measurements of the London penetration depth \cite{K2Cr3As3Pen}. We reported measurements of the London penetration depth on polycrystalline samples of K$_2$Cr$_3$As$_3$, where the samples with the sharpest superconducting transitions display linear behavior below about 1.4~K  \cite{K2Cr3As3Pen}, which is strong evidence for the presence of line nodes in the superconducting gap. The samples are also extremely air sensitive and measurements of samples with broadened superconducting transitions show $\Delta\lambda(T)\sim T^n$ dependence with $n~>~1$, which emphasizes the necessity of studying high quality samples which have not partially decomposed.

A number of interesting properties of the normal state of K$_2$Cr$_3$As$_3$ have been reported such as a linear temperature dependence of the electrical resistivity between 7 and 300~K \cite{Bao2015superconductivity}, and a temperature dependence of the spin relaxation rate from NMR measurements consistent with that of a Tomonaga-Luttinger liquid \cite{K2Cr3As3NMR}. The specific heat shows a bulk superconducting transition and gives a large value of the Sommerfeld coefficient $\gamma$ of around 70~-~75~mJ/mol~K$^2$ \cite{Bao2015superconductivity}, which indicates enhanced electronic correlations. Upon increasing the size of the alkaline metal cation $A$, there is a significant increase in the lattice parameter $a$ and a reduction of $T_c$. This would suggest that increasing the interchain distance suppresses superconductivity, but seemingly the opposite result is obtained from measurements under pressure, where increasing the pressure causes a decrease in the $T_c$ of K$_2$Cr$_3$As$_3$ and Rb$_2$Cr$_3$As$_3$ \cite{K2Cr3As3Press}. This suggests that the superconductivity in $A_2$Cr$_3$As$_3$ is sensitive to distortions of the  [(Cr$_3$As$_3$)$^{2-}]^\infty$ tubes \cite{Crtubes}.

It is therefore of interest to establish to what extent the superconductivity can be regarded as q1D and what is the nature of the pairing state in these compounds. The calculated electronic structure of  K$_2$Cr$_3$As$_3$ is complex, with two one-dimensional Fermi surface sheets ($\alpha$ and $\beta$) and one anisotropic three dimensional sheet ($\gamma$) \cite{K2Cr3As3Elec}. Theoretical studies using a three-band tight-binding model predict the dominant pairing states to be spin triplet, namely the $p_z$ pairing state for small $J/U$ and an $f$-wave state for large $J/U$ \cite{TheorYiZhou,Theor3}, where $U$ is the intra-orbital Coulomb repulsion and $J$ is the Hund’s coupling. The $p_z$ state has a nodal plane at $k_z~=~0$, although the pairing amplitude is strongest on one of the q1D sheets, which are at fixed values of $k_z$ and the gap is fully open and isotropic on these surfaces. The $f$-wave pairing is strongest on the 3D $\gamma$ sheet, where there are line nodes in the $ab$~plane. Initial measurements of the upper critical field ($H_{c2}$) showed large values with a small anisotropy \cite{Tai2015anisotropic}. The extrapolated value of $H_{c2}(0)$ was estimated to be about 31.2~T, which is much greater than the Pauli limiting value for weak coupling BCS superconductors of $1.84T_c\sim11.2$~T, which was taken as evidence for triplet superconductivity. However, recent experiments performed at higher applied magnetic fields \cite{Hc2HighB} demonstrate that a larger anisotropy develops at lower temperatures, which may indicate that contrary to the predictions of spin triplet superconductivity, paramagnetic limiting is present for fields applied along the chains but is absent for fields applied perpendicularly.

Linear behavior of the penetration has also been predicted to occur in measurements of polycrystalline superconductors as a result of Josephson coupling between different grains \cite{JosGran}. Muon spin rotation measurements of  K$_2$Cr$_3$As$_3$ were unable to discriminate between nodal superconductivity and a fully gapped model with a smaller gap than the BCS value \cite{K2Cr3As3MuSR}.  Therefore to confirm the intrinsic nature of the behavior of the London penetration depth, in this work we report measurements of single crystals using a tunnel diode oscillator (TDO) based technique.  A downturn in the frequency shift ($\Delta f$) is consistently observed in all measured samples at around 0.9-1.2~K. However, above the downturn there is good agreement with the polycrystalline data and linear behavior extends from about 1.4~K to the onset of the downturn. Although further work is required to improve the quality of single crystal samples and to determine the origin of the downturn, these results indicate that the linear behavior observed in polycrystalline samples is not due to Josephson coupling between grains, but is an intrinsic property of the compound.

\section{Experimental details}

Polycrystalline samples of  K$_2$Cr$_3$As$_3$ and Rb$_2$Cr$_3$As$_3$ were prepared by a solid state reaction described in Refs.\cite{Bao2015superconductivity} and \cite{Tang2015unconventional}. Single crystals of K$_2$Cr$_3$As$_3$  were synthesized by spontaneous nucleation using a self-flux of As-K \cite{Bao2015superconductivity}. The London penetration depth was precisely measured using a self-induced tunnel diode oscillator based method down to about 0.4~K in a $^3$He cryostat.  Since the samples are very air sensitive and decompose very quickly upon being exposed, they were always handled in an Argon glove box and were kept in Apiezon N grease when being measured. During the measurements, a very small alternating field of about 20~mOe is applied, which is much smaller than the lower critical field. As a result, the sample is always in the Meissner state so that the measured frequency shift $\Delta f$  is proportional to the London penetration depth, which is given by $\Delta\lambda(T)=G\Delta f(T)$, where the calibration constant $G$  is solely dependnt on the sample and coil geometry\cite{Prozorov2000meissner}.

\section{Results and discussions}

\begin{figure}[tb]
\begin{center}
  \includegraphics[width=0.75\columnwidth]{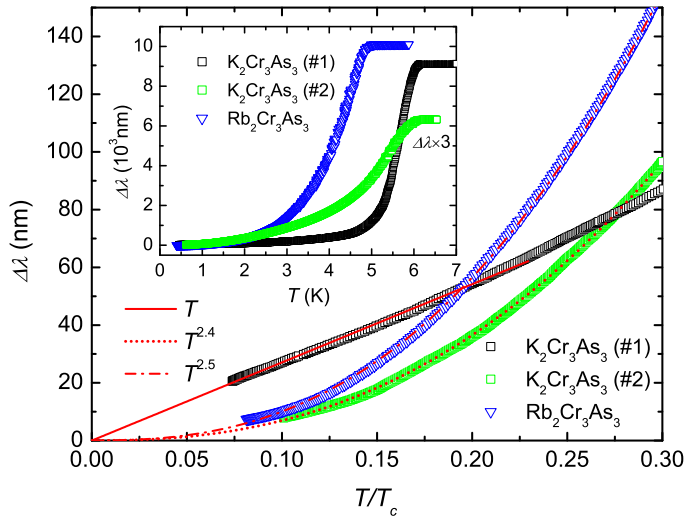}
\end{center}
	\caption{(a) Temperature dependence of the London penetration depth $\Delta\lambda(T)=\lambda(T)-\lambda(0)$  at low temperatures for polycrystalline K$_2$Cr$_3$As$_3$ (No.~1 and No.2) \cite{K2Cr3As3Pen} and Rb$_2$Cr$_3$As$_3$ . The solid line shows the linear behavior of sample No.~1 below 1.4~K, while the dotted and dashed lines show the  $\Delta\lambda(T)\sim T^n$ behavior of the other two samples with $n~=~2.4$ and $2.5$ for sample No.~2 and  Rb$_2$Cr$_3$As$_3$ respectively. The inset shows  $\Delta\lambda(T)$ measured from 0.4~K to above $T_c$ for the three samples, where the transition for the air exposed  sample No.~2 is considerably broader than No.~1 and the data have been scaled by a factor of three.}
   \label{figure2}
\end{figure}

Figure~\ref{figure2} displays the temperature dependence of the change of London penetration depth $\Delta\lambda(T)$ from about 7~K to 0.4~K for two samples of polycrystalline K$_2$Cr$_3$As$_3$, which we reported previously \cite{K2Cr3As3Pen}. A sharp transition is observed for sample No.~1 with an onset temperature of 6.1~K, which is consistent with previous measurements \cite{Bao2015superconductivity}. As a result of the air sensitivity of the samples, it was not possible to measure both $\Delta\lambda(T)$, and other physical properties of the  same sample to check the sample quality,  but an indication of the sample quality is given by the sharpness of the superconducting transition in the measurements of $\Delta\lambda(T)$. For samples which are properly handled during the experimental preparation, a sharp superconducting transition is typically observed and the physical properties are highly reproducible. However, when the samples have had a greater exposure to air, the superconducting transition is significantly broader, as shown for example by sample No.~2.   A clear linear temperature dependence of $\Delta\lambda (T)$ is observed from the base temperature of 0.4~K to about 1.4~K in sample No.~1, which is unlike the behavior of fully gapped superconductors, where there is an exponential dependence and saturation at low temperatures. The linear behavior was found to be reproducible in another sample with a sharp superconducting transition \cite{K2Cr3As3Pen}. Similar behavior is observed in cuprate and some heavy fermion superconductors, which implies line nodes in the superconducting gap \cite{Jacobs1995plane,Bonalde2005evidence}. In Ref.~\cite{K2Cr3As3Pen}, we fitted the superfluid density with several models with different gap structures, but we were unable to discriminate between the various models with line nodes. Since the crystal structure of  K$_2$Cr$_3$As$_3$ is noncentrosymmetric and there is evidence of Pauli paramagnetic limiting of $H_{c2}$ only along some crystallographic directions \cite{Hc2HighB}, the line nodes may arise due to the mixture of spin singlet and triplet pairing expected to occur in superconductors lacking inversion symmetry \cite{BauerNCS}. In addition, we cannot rule out the possibility of 
multi-band superconductivity in  K$_2$Cr$_3$As$_3$ and if this is the case, our results imply that at least one of the gaps has line nodes. Measurements of a sample of  Rb$_2$Cr$_3$As$_3$ are also shown in Fig.~\ref{figure2}. As displayed in the inset, there is an onset of superconductivity at around 4.9~K, which is consistent with previous reports \cite{Tang2015unconventional}. The superconducting transition is broad, which may indicate that the sample has partially reacted with air. The data do not show exponential behavior at low temperatures but a $\Delta\lambda(T)\sim T^n$ dependence, with $n~=~2.5$ and the value of $n$ was also found to be sample dependent. This is very similar behavior to that of the  K$_2$Cr$_3$As$_3$ samples which have had a greater exposure to air, as shown by sample No.~2, where $n~=~2.4$. These results suggest that a similar scenario may exist for  K$_2$Cr$_3$As$_3$ and  Rb$_2$Cr$_3$As$_3$, and measurements of higher quality samples of   Rb$_2$Cr$_3$As$_3$ are required to determine whether $\Delta\lambda (T)$ is linear at low temperatures.

\begin{figure}[tb]
\begin{center}
  \includegraphics[width=0.75\columnwidth]{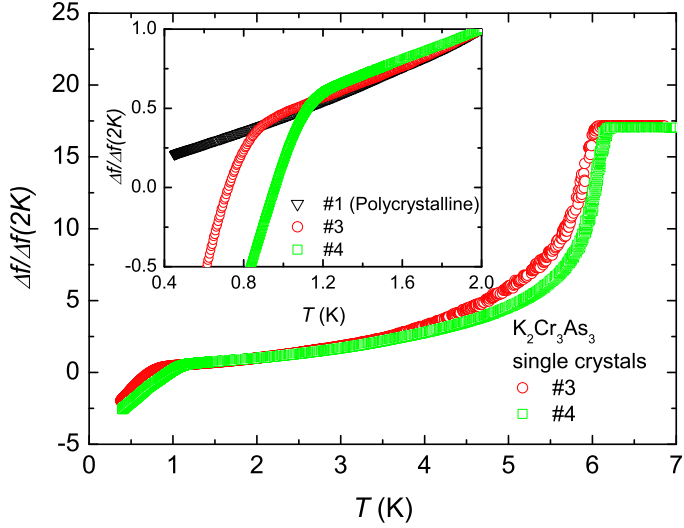}
\end{center}
\caption{The frequency shift $\Delta f(T)= f(T)- f(0)$, as a function of temperature from 0.4~K to 7~K for two single crystals of K$_2$Cr$_3$As$_3$, one synthesized at Zhejiang University (No.~3), the other at the Institute of Physics in Beijing (No.~4). Note that $f(0)$ has been determined by extrapolating the behavior above the low temperature downturn to $T~=~0$. The data have been normalized by the 2~K value. The inset shows the low temperature behavior below 2~K for the two single crystal samples and a polycrystalline sample (No. 1).}
\label{figure3}
\end{figure}

Due to the possibility of a linear contribution to $\Delta\lambda(T)$ in polycrystalline samples from Josephson coupling \cite{JosGran}, measurements were performed on several single crystals prepared by two different research groups so that any intergrain coupling is greatly reduced. Measurements of two samples, one synthesized at Zhejiang University (No. 3) and another from the Institute of Physics in Beijing (No. 4) are shown in Fig.~\ref{figure3}. The calibration constant $G$ for the single crystal samples can not be reliably estimated due to the very small needle-like shapes, instead of the geometry of a regular square plate. As a result, we normalized the experimental data to the value at 2~K. Since the samples were mixed with grease, the relative orientation of the applied field and the needle axis was not determined. The sharp decrease at about 6.1~K corresponds to the onset of superconductivity which is consistent with the results from polycrystalline material. 

It can be seen that for both samples there is no evidence for BCS-like saturation of $\Delta f(T)$ and below 1.4~K, there is a linear decrease, much like the measurements of polycrystalline samples. It should be noted that for an isotropic, weakly coupled BCS superconductor, the signal is expected to saturate at approximately 0.2$T_c$ and since the jump in the specific heat $\Delta C/\gamma T_c\sim2.2$ \cite{Tai2015anisotropic} is significantly larger than the BCS weak coupling value of 1.43, if the system is an isotropic BCS superconductor, the saturation would be expected to occur at higher temperatures still. However, at lower temperatures a downturn in $\Delta f(T)$ is consistently observed in all measured samples with an onset between 0.9 and 1.2~K. Upon further lowering the temperature, the signal continues to decrease steadily down to the base temperature of about 0.4~K. The size of the drop in $\Delta f(T)$ below this anomaly compared to that below the superconducting transition is also sample dependent. In the inset of Fig.~\ref{figure3}, the normalized $\Delta f(T)$ at low temperatures is compared for the single crystal and polycrystalline measurements, which shows that above the anomaly similar linear behavior is observed. Although the normalized values of $\Delta f(T)$ agree at low temperatures above the downturn, this does not necessarily indicate that the superfluid density of single crystals is the same as that of the polycrystalline samples reported in Ref.~\cite{K2Cr3As3Pen}. For polycrystalline samples, the superfluid density is an average over all field orientations, while for a single crystal sample, only one field direction is measured. If the gap structure is anisotropic the superfluid density would be different for single crystal and polycrystalline samples, but if there are line nodes in the superconducting gap, both may show a linear temperature dependence at low temperatures.

We are not able to definitively determine the origin of the low temperature downturn in the single crystal samples. It is unlikely that this anomaly arises due to decomposition from exposure to air. As shown by sample No.~2 in  Fig.~\ref{figure2}, this would be expected to lead to a broader superconducting transition and instead of linear behavior,  $\Delta\lambda(T)\sim T^n$ with $n~>~1$ is observed \cite{K2Cr3As3Pen}. The single crystals display sharp superconducting transitions, similar to that observed in sample No.~1. One possibility is that the downturn occurs due to small amounts of superconducting impurites formed from the excess As-K flux. Although there are several binary phases of As-K \cite{KAsbin}, the properties of many of these have not been characterized below 2~K and we were not able to identify a matching superconducting phase in the literature. A downturn in $\Delta f(T)$ can also arise from the presence of a magnetic transition or due to the proximity effect from a thin layer of normal material on the sample surface \cite{TDOAFM,Prozorov2006magnetic}.

\section{Summary}

To summarize, we have measured the frequency shift  $\Delta f(T)$ of single crystals of K$_2$Cr$_3$As$_3$,  which is proportional to the change in the London penetration depth. At low temperatures there is no evidence for the saturation behavior expected for fully gapped superconductors and below about 1.4~K, a region of linear behavior is observed, in agreement with measurements of polycrystalline samples. This is a strong indication that the linear temperature dependence measured on polycrystalline samples is not due to Josephson coupling but from line nodes in the superconducting gap, indicating that the pairing state is unconventional. However, in all single crystal samples, a downturn in  $\Delta f(T)$ is observed at low temperatures,  with an onset in the range of 0.9 to 1.2~K. Further work is required to improve the quality of single crystal samples and to understand the cause of the low temperature downturn. We also reported measurements of $\Delta\lambda(T)$ for polycrystalline samples of Rb$_2$Cr$_3$As$_3$, which display a broad superconducting transition and a $\Delta\lambda(T)\sim T^n$ dependence at low temperatures with $n~>~1$. This is similar behavior to that observed in samples of K$_2$Cr$_3$As$_3$ which have partially decomposed due to exposure to air, but measurements of high quality polycrystalline and single crystal samples are required to determine whether there is the same linear temperature dependence at low temperatures.

\section*{Acknowledgments}
This work was supported by the National Basic Research Program of China (No 2011CBA00103), the National Natural Science Foundation of China (No.11474251 and No.11174245) and the Fundamental Research Funds for the Central Universities.

\end{document}